# A Novel Approach to Radiometric Identification

Raoul NIGMATULLIN[a], Semyon DOROKHIN[b] and Alexander IVCHENKO[b, 1]
[a]*Kazan National Research Technical University named after A.N. Tupolev, (KNRTU-KAI)*
[b]*Moscow Institute of Physics and Technology (MIPT)*

**Abstract.** This paper demonstrates that highly accurate radiometric identification is possible using CAPoNeF feature engineering method. We tested basic ML classification algorithms on experimental data gathered by SDR. The statistical and correlational properties of suggested features were analyzed first with the help of Point Biserial and Pearson Correlation Coefficients and then using P-values. The most relevant features were highlighted. Random Forest provided 99% accuracy. We give LIME description of model behavior. It turns out that even if the dimension of the feature space is reduced to 3, it is still possible to classify devices with 99% accuracy.

## 1. Introduction

The security of a wireless network can be significantly improved if there is a device identification algorithm. Review [1] provides sufficient examples, such as authentication & flooding attack and Access Point forgery. The features for identification can be acquired from application, transport, MAC and physical layers [1]. Application and transport layer do not provide a large number of features. Therefore, they are quite rarely chosen in device identification. MAC layer provides vendor-specific features that are based on implementation of underspecified standard details. If one obtains the technical details of that realization, a new forgery can be committed.

Physical layer features can be described as channel-specific or device-specific. Channel-specific features ought to be neglected in mobile devices identification, since such devices change their location thus altering features' value. Device-specific features are based on minor imperfections in electronic components. These imperfections constitute a unique set of features thus allowing reliant identification. RF identification by device-specific features is called radiometric identification. A brief overview of radiometric identification methods is presented in Table 1.

Radiometric identification can be performed in waveform or modulation domain. A classical modulation domain algorithm was developed by Brik et al. [2]. The researchers analyzed modulation errors to obtain features. The method was tested on 138 network interface cards and proved to be reliable. However, such methods require knowledge of modulation scheme and imply proper synchronization and channel estimation, which may be difficult to achieve.

Among the methods operating in waveform-domain, there is a class of algorithms based on transient analysis. Barbeau et al. [3] generated features using wavelet transform. This method requires an additional acquisition algorithm to track the start of transient process. Adam Polak et al. [4] resorted to a more fundamental approach. The researches modeled DAC and amplifier nonlinearities and used the models' coefficients for further classification. Though this method illustrates the nature of radiometric fingerprint, the amplifiers need to be analyzed on evaluation boards.Several research groups tried to use features based on high-order statistics. The general approach is to split the signal into several regions of interest (ROI) and to calculate variance, skewness and kurtosis for amplitude, frequency and phase of each ROI [5], [6]. Trevor Bihl et al. [6] proved that statistical features extracted from the signal phase play the major role. However, such approach has low average true verification rate.Recently there were several attempts to solve the task of radiometric identification with neural networks. O'Shea et al. [7] used high-order statistics to generate features for CNN. This resulted in quite a good classification at high signal-to-noise ratio (SNR).

**Table 1.** Comparison of existing radiometric identification methods

| Authors | Features' nature | Advantages | Disadvantages |
|---|---|---|---|
| Brik et al. [2] | Modulation accuracy | Tested on large (138) number of devices | Requires demodulation |
| Barbeau et al. [3] | Transient wavelet coefficients | Tested on 10 devices | Requires acquisition of transient |
| Polak et al. [4] | Nonlinearity model | Error probability expressed analytically | Additional equipment needed |
| Bihl et al. [6] | High-order statistics | Importance of features analyzed | Low average true verification rate (71%) |
| O'Shea et al. [7] | Convolutional neural network | High accuracy for different constellations | High computational complexity |

This paper focuses on feature engineering for radiometric identification. A novel method is proposed based on the recent paper [8]. The new parameters allow to achieve high accuracy (97%) without the need for computationally complex feature generation and complicated classification algorithms. The designed algorithm is tested on experimental data. Since it is the first attempt to apply methods described in [8] to radiometric identification, we consider a simple task of binary classification, with one receiver and only two transmitters.We subtract etalon signal from the received one and analyze only the phase of the resulting signal, as it was shown to be more relevant [6].

The rest of the paper is organized as follows: section 1 provides the detailed description of the features used for classification, section 2 describes the experimental setup and the process of dataset generation, followed by section 3, where the results are demonstrated and important comments are given.

## 2. The Proposed Features

In this section we outline the basic principles of the CAPoNeF (Comparative Analysis of the Positive and Negative Fluctuations) method. Usually, in the conventional statistics only the mean value and standard deviation are frequently used for description of trendless sequences (TLS). In the recent paper [8] it was shown that a new set of quantitative parameters can be suggested for quantitative description of various TLS. These parameters have different sensitivity to the hidden random factors. More important, they are free from treatment errors and model assumptions. In this paper, we choose the following parameters to characterize a specific competition between positive and negative fluctuations of the given rectangle matrix $y_j(m)$ ($j = 1, 2, \ldots N$ – the number of data points/matrix rows; $m = 1, 2, \ldots M$ the number of columns/successive measurements). For the selected column of the given rectangle matrix (we omit the column index $m$ for simplicity) the chosen quantitative parameters are the following:

$P_1$ = arithmetic mean($y$) coincides with mean value of the given TLS.

$P_2 = \max(Yup) - \min(Ydn)$ – the range between positive and negative fluctuations.

$P_3 = \max(Yup) - |\min(Ydn)|$ – the relative fluctuation intensity.

$P_4 = \max(Jup) - \min(Jdn)$ – the maximal cumulative intensity between positive/negative fluctuations. ($Jup/Jdn$ corresponds to the integrals taken from initial TLS(s) $y_j(m)$).

$P_5 = [\max(y)-\text{mean}(y)]/[\text{mean}(y)-\min(y)]$ – measure of asymmetry. If $P_5 = 1$, then the TLS can be considered as "ideally" symmetrical sequence relatively its mean value.

$P_6 = \text{last}(x_{up}) - \text{last}(x_{dn})$ – this parameter serves as an asymmetrical measure in the horizontal direction.

$P_7 = \max(Js)$ - maximal value of the bell-like curve formed from sequence of the range amplitudes (SRA). This SRA, when the amplitudes ($y_1 > y_2 > \ldots > y_N$) are organized in the descending order, is integrated and forms the bell-like curve. The maximal value separates the positive/negative branches of the ordered fluctuations.

$P_8 = \text{Range}[J(y_n)]$, $y_n = (y - \langle y \rangle)/\text{Range}(y)$, $\text{Range}(y_n)=1$, where $\text{Range}(f)=\max(f)-\min(f)$. This parameter is also important, because it allows to compare the integral ranges taken from equivalent/normalized columns having the unit range value.

We would like to emphasize that any TLS $Dy_j = y_j - \langle y \rangle$ oscillates near zero value. We noticed that the distribution of the roots numbers is generally described by the segment of the straight line $R_k \cong a \cdot k + b$, where $a$ and $b$ are regression parameters. This observation allows to find the mean oscillation frequency and phase from the equation $\cos(\langle \omega \rangle R_k - \Phi) = 0$ or $\langle \omega \rangle R_k - \Phi = \pi/2 + \pi \cdot k$. It means that one can add two additional parameters invariant to scaling, mean frequency $\langle \omega \rangle$ and phase $\Phi$:

$P_9 = \langle \omega \rangle$.

$P_{10} = \Phi$.

## 3. Experimental Setup

The data was collected using Adalm Pluto software-defined radio (SDR). One SDR was used as a receiver, while the two other devices were used as transmitters. The signal of L = 1024 samples was broadcasted from one transmitter at a time, repeated in a cyclic fashion. The signal was also known at the receiver side, which allowed to perform time-domain tracking and compensate for sampling clock offset. The synchronized signal was normalized and the etalon signal was subtracted from every group of L samples. Thus, we obtained the error signal. Its phase was used for further feature generation.

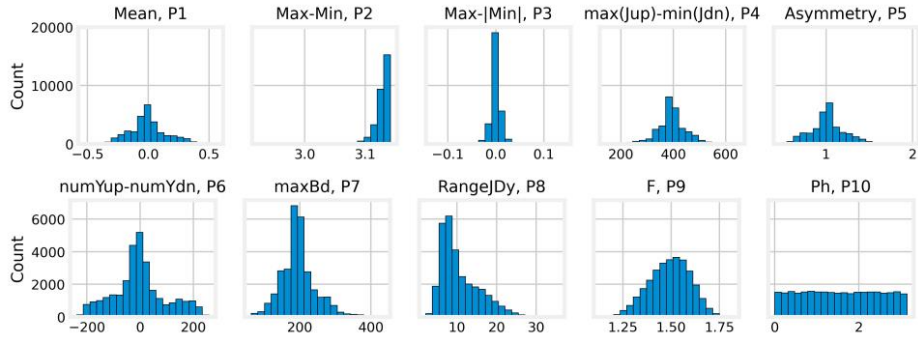

**Figure 1**. Hystograms of the features

We assume that the choice of the signal itself does not affect the statistical properties of the error signal significantly. Therefore we felt free to use "trans-noise" – a noise-like signal, generated from the first and the second L digits of transcendental number π. Every digit was mapped to certain amplitude, so that 0 and 9 corresponds to minimal and maximal DAC level.

Experimental data was collected in the following setup. First, the main lobes of TX and RX antennas radiation patterns where facing each other, the distance between the devices changed from 1 to 5 meters. Then the distance between the devices was fixed (2 meters), but the angle between the main lobes changed from –π/2 to π/2 with the step of π/6. The resulting dataset contains 30000 labeled entries.

## 4. Results and Discussion

Prior to classification, we analyzed statistical properties of the proposed features. Histograms were calculated for every feature to study their distributions (see Figure 1). After that we studied correlation properties. We calculated the Pearson Correlation Coefficient (PCC) for every pair of features. As it can be seen in Figure 2, there is high correlation between the parameters $P_1, P_5$ and $P_6$ (PCC reaches 0.96). This peculiarity should be taken into consideration. We will show that eliminating these parameters improves the accuracy. We also calculated Point Biserial Correlation Coefficient (PBCC) and P-value for every feature. The P-value of the features' significance level is

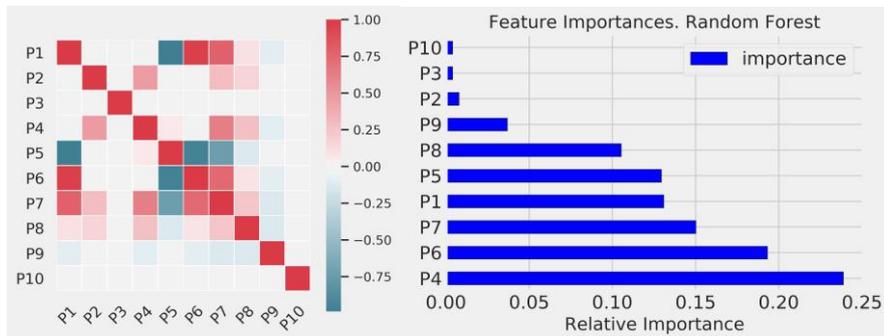

**Figure 2**. Heatmap (left) and feature importances (right)

0.05. If the P-value level is less than 0.05, the parameter is considered statistically significant. Table 2 contains PBCCs and P-values for the dataset. The most interesting parameters are $P_9$, $P_8$ and $P_2$. The parameters $P_4$, $P_3$, $P_{10}$ have P-values higher than 0.05.

**Table 2.** Point Biserial Correlation Coefficients and P-values

| Parameter | PBCC | P-value |
|---|---|---|
| $<\omega>$, $P_9$ | **0.3025** | 0.0 |
| Range$[J(y_n)]$, $P_8$ | **0.4293** | 0.0 |
| max($Yup$)– min($Ydn$), $P_2$ | **0.1227** | 0.0 |
| max($Jup$) – min($Jdn$), $P_4$ | 0.0053 | 0.3602 |
| max ($Yup$) – |min($Ydn$)|, $P_3$ | -0.0008 | 0.8839 |
| $\Phi$, $P_{10}$ | -0.0003 | 0.9517 |
| max($Js$), $P_7$ | -0.0173 | 0.0027 |
| last($x_{up}$)-last($x_{dn}$), $P_6$ | -0.0163 | 0.0047 |
| mean($y$), $P_1$ | -0.0179 | 0.002 |
| [max($y$)-mean($y$)]/[mean($y$)-min($y$)], $P_5$ | -0.0662 | 0.0 |

We compared basic ML classification algorithms. Among all classifiers, such as Linear Regression, Decision Tree and KNN, Random Forest showed the best results with the accuracy of 0.99.

*4.1. Feature Importances and LIME*

Feature importances show the relevance of every parameter. They are defined as the total decrease in node impurity, weighted by the probability of reaching that node, averaged all over the trees of the ensemble. Probability of reaching the given node is approximated by the proportion of samples reaching that node [9]. The most predictive features are P3, P6, and P7. Locally Interpretable Model-agnostic Explanations (LIME) [10] shows how the model makes decisions by approximating the area around the forecast using a linear model. We use it to explain the reason for wrong predictions and proper behavior (Figure 3).

The plot from LIME shows the contribution to the final forecast of each of the example parameters. In case of correct classification (Figure 3, left) P1, P2, P8, P5, P10, P4, P6 indicate the transmitter #0 and increase the accuracy, while P9, P3, P7 indicates transmitter #1 and has a negative impact. In case of wrong prediction P1, P3, P6 indicate the transmitter #0, but others indicate transmitter #1 (Figure 3, right).

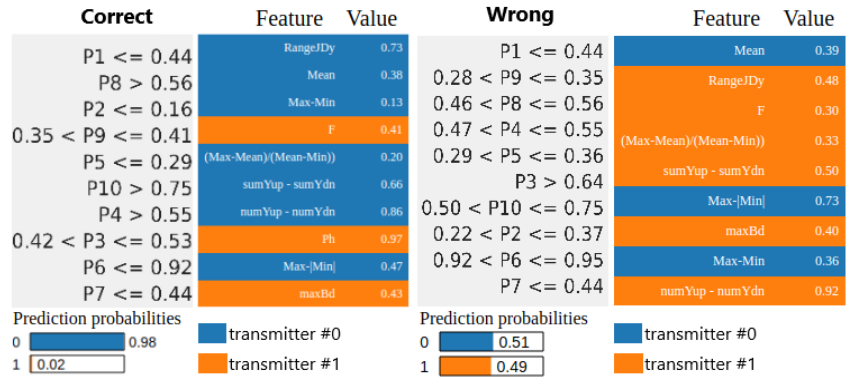

**Figure 3**. LIME description of wrong (right) and correct (left) prediction.

*4.2. Feature Selection*

According to Section 4, we have provided the experiments on reducing the feature dimension. We focused on Random Forest algorithm, that gives us the best results on full dataset and applied 4-folds stratification and grind search of optimal hyperparamers on 40 iterations. Removing parameters P5 and P6 gives approximately the same accuracy (0.9863 versus 0.9888). Training only on the 3 most important parameters P4, P5 or P7, P8 gives 0.9858 accuracy. We find them applying LIME analysis, Point Biserial correlation coefficient and Random Forest feature importances. We assume that other features will contribute more in multiclass case under more complex conditions. It is difficult to compare with existing methods since the results depend on the dataset and the task. However, the dataset is available online [11] for further comparative analysis.

*4.3. Conclusion*

This paper demonstrates the application of the CAPoNeF method for radiometric identification (binary classification). We collected data using SDR with various distances and angles between the devices. We subtracted etalon signal from the received one and extracted 10 parameters from the phase of the resulting signal. The parameters were further analyzed using the ML. P2 (the range between positive and negative fluctuations), P8 (the range of the integral of the normalized signal) and P9 (mean frequency) have the highest predictive ability. Classification task was solved: Random Forest classifier showed 99% accuracy.

In further research, we are planning to solve multiclass classification task. Current research did not take into account possible fading. Additional study should be carried out to determine whether the proposed method is robust to fading environment.